\documentclass{article}
\usepackage[xdvi]{epsfig}
\begin{document}
\title{The Myth of the Twin Paradox}
\author{E. Fischer\footnote{e-mail: e.fischer.stolberg@t-online.de}}
\date{Auf der Hoehe 82, 52223 Stolberg, Germany}
\maketitle
\begin{abstract}
One of the most discussed peculiarities of Einstein's theory of relativity is
the twin paradox, the fact that the time between two events in space-time
appears to depend on the path between these events. We show that this time
discrepancy results only from faulty assumptions in the transition from one
reference system to another. The twin paradox does not exist. But the Lorentz
invariance of the theory has strong consequences, if we assume that it is
valid not only locally, but also on cosmic scale.
\end{abstract}
\section{Introduction}
No other development in physics has changed our view of the world more than
the theory of relativity, introduced by Einstein in 1905. Based on the idea
that motion of a body cannot be defined absolutely, but only relative with
respect to others, he concluded that the laws of physics should look alike in
any two reference systems, moving with respect to each other with constant
velocity. Together with the experimentally confirmed fact that the speed of
light is independent of the reference frame, this led to the conclusion that
time cannot be an absolute quantity, but that time and space constitute a
4-dimensional unit. Temporal and spatial distances both must depend on the
reference system.

The equivalence of all inertial reference systems requires that linear motion
in one system translates into linear motion in another system. Thus the
relation between the coordinates $(x,y,z,t)$ in one system and those in
another system $(x',y',z',t')$, moving with velocity $v$ in the direction
$x$, must be linear. These conditions uniquely define the transformation
equations (the Lorentz transformation)
\begin{equation}
\label{lor}
x'=\gamma (x-\beta t)\qquad y'=y \qquad z'=z \qquad t'=\gamma(-\beta x+t)
\end{equation}
where the time variable is calibrated to an equivalent length by the speed of
light $c$ $(ct\Rightarrow t)$. $\beta$ is the relative velocity as a fraction
of $c$: $\beta=v/c$ and $\gamma =1/\sqrt{1-\beta^2}$. The two systems are
synchronised by the condition that at $x=t=0$ we have $x'=t'=0$.

The fact that the time scale changes with distance appears somewhat strange
to our intuition, but this can be ascribed to the fact that in every days
life we are accustomed only to velocities, which are much less than $c$, so
that relativistic effects are negligible. On the other hand the lack of
imagination has led us to believe in mathematically derived consequences
which can scarcely be proved by experiments.

One of the most discussed consequences of the theory of relativity is the so
called twin paradox, which dates back to first decade after the invention,
but is discussed in numerous scientific papers still today. Basis of the twin
paradox is the dilatation of time, the fact that moving clocks are slowed
down, when observed from the rest system. Thus a clock moving with a
considerable fraction of $c$ measures a shorter time to reach a distant
target than a clock at rest.

The twin paradox is frequently told with the following story: There are twins
Alice and Bob. Alice decides to make a journey to a distant star in a
spaceship capable of moving at a considerable fraction of $c$. When she has
reached the star, she goes back at the same speed. According to time
dilatation a clock moving with her and consequently also Alice herself ages
more slowly than her sibling at home, so that, when she comes back, she finds
Bob as an old man, while for herself the journey has taken only a few years.

In this paper we will show that this interpretation of relativity is
incorrect and that the twin paradox does not exist at all. We will try to
show up, where the mathematical flaws come in, and how they can be corrected.
Subsequently we will consider some real effects of relativity, especially
with respect to the consequences of Lorentz invariance in accelerated
systems, as they are discussed in general relativity.

\section{Invariants of the Lorentz transformation}
The Lorentz transformation describes, how spatial and temporal distances
between events change, when observed from different reference systems, which
are in relative motion with respect to each other, with the additional
condition that the speed of light is independent of the reference system. The
equation $x=t$ transforms into $x'=t'$ (time calibrated by $c$ as in the last
section), independent of the direction of the relative motion of the systems.

This property can be read immediately from the transformation equation
eq.(\ref{lor}), defining the space-time distance between two events by
\begin{equation}
\Delta s=\sqrt{\Delta x^2+\Delta y^2+\Delta z^2-\Delta t^2}.
\end{equation}
or, if the argument of the square root is negative, the proper time
\begin{equation}
\Delta \tau=\sqrt{\Delta t^2-\Delta x^2-\Delta y^2-\Delta z^2}.
\end{equation}
Leaving off the $y$ and $z$ coordinates for simplicity, as these are not
affected by the transformation, from eq.(\ref{lor}) we get
\begin{equation}
\label{del}
\Delta \tau=\sqrt{\Delta t^2-\Delta x^2}=\sqrt{\Delta t'^2-\Delta x'^2},
\end{equation}
which reduces to $\Delta \tau=0$ in the case of light. This is the well known
fact that the world lines defined by $\Delta \tau=0$ constitute the limits of
the region, which can be causally connected to some space-time event.
Invariance of the condition $\Delta \tau=0$ means that causality is not
affected by any Lorentz transformation.

But with respect to the twin paradox it is more important that the quantity
$\Delta \tau$ is generally invariant under Lorentz transformations, not only
in the case $\Delta \tau=0$. The space-time distance between two events is a
uniquely defined quantity, independent of the reference system. If two events
take place at the same physical location, we must assume that their spatial
distance is zero in every reference system. As their space-time distance is
uniquely defined, too, the necessary consequence according to eq(\ref{del})
is that the temporal distance is also uniquely defined. It must be
independent of the choice of the reference system.

For Alice and Bob there are two events, where they are physically at the same
spatial position: The first one, when Alice leaves the earth with her
spaceship, and the second one, when she returns from her journey. The
space-time distance between these events is uniquely defined and thus also
the time interval between the events. There is no room for any discrepancy in
the aging of Alice and Bob.

Switching from one reference system to another may change the local time
scale during her journey, but changes of the reference system cannot change
the underlying physics. In the next section we will try to find out, where
the pitfalls are, which lead to positive results for the twin paradox.

\section{False solutions}
False solutions of the twin paradox date back to the very beginning of the
theory of relativity. Even Einstein himself has mentioned time discrepancies
as a peculiar result of his theory \cite{einstein} and the story of the
traveling twins was introduced by Langevin as early as 1911 \cite{langevin}.
Numerous papers on the topic have been published since then, but most of them
give more or less sophisticated explanations, why the twins age differently,
but the existence of the effect is scarcely disputed.

The most simple argument in favour of a different aging runs as follows. The
clock in the moving spaceship is slowed down with respect to a clock on earth
by a factor $1/\gamma =\sqrt{1-\beta^2}$. Thus the time to reach the point of
return is reduced just by this factor. As the return journey is symmetric
with respect to time, the total time is also reduced by this factor compared
to the elapsed time on earth.

This reasoning contains several errors, however. The first one is that it
regards the spatial geometry, which is fixed in the rest frame of the earth,
as fixed also in the comoving frame and only considers the change of the time
scale. The second is that at the return point there is a further change of
the reference system. While the Lorentz transformation between the earth and
the spaceship has been synchronised at the starting point, there is no such
synchronisation at the return point. That means that the zero point of the
time scale is altered. Closely related to the synchronisation problem is the
fact that the Lorentz transformation does not conserve simultaneity.
Simultaneous events at different space points do not remain simultaneous with
changes to a moving coordinate system.

It should be stressed here that a proper time interval is not the change of
some scalar property between two events, but it is the equivalent in a
pseudo-Euclidean metric to the vector length in Euclidean geometry. Thus
adding proper time intervals, measured in different reference frames, must be
done by the rules of vector algebra and not like the addition of scalar data.
Change to a moving reference system is analog to rotation in Euclidean space.

The problem can easily be visualised in Euclidean geometry (see fig.1). We
consider a set of transformations, consisting of a rotation of the $(x,y)$
coordinate system by some angle $\vartheta$ about the zero-point to
$(x',y')$, subsequent shifting the coordinate system in $x'$-direction to
$(x'',y'')$, rotating the system back by $-\vartheta$ to $(x''',y''')$ and
finally shift it back in direction of $x'''$ to $(x'''',y'''')$, so that
$x''''=x$. At the end $y''''$ will not be equal to y. But this operation does
of course not change any distance between points in the $(x,y)$ plane.

\begin{figure}[hbt]\epsfig{file=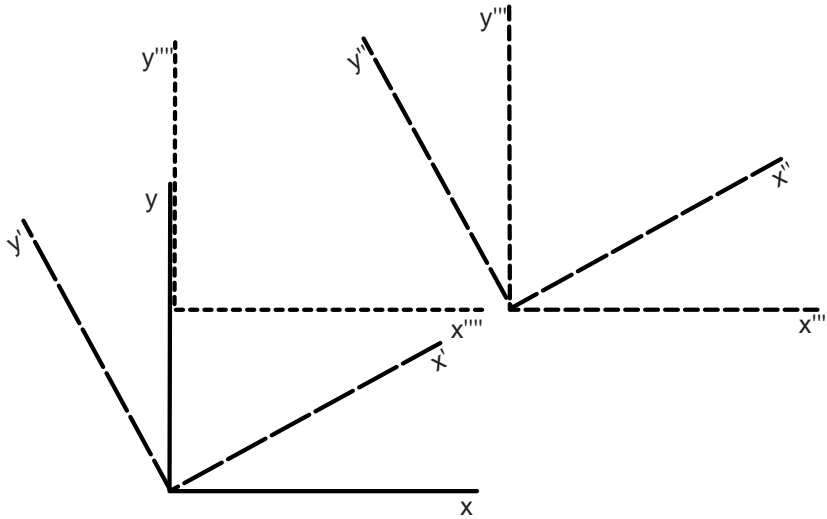,width=11cm}
\caption{Coordinate transformations in Euclidean space: rotation, shift
along $x'$ axis, back rotation, back shift along $x'''$ axis}
\end{figure}

It is the fact that we have used rotations with different centers, which
leads to a shift of the y coordinate. In just the similar way the age shift
of the twin paradox results from the fact that the reference frame is changed
twice, first at the starting point, and then a second time at the point of
return, but now with a different center of 'rotation'. It is this change of
velocity at the return point, which causes the supposed age shift. To reach
the velocity of the new comoving reference frame, the spaceship has to be
accelerated. During the acceleration phase the proper time changes
continuously. An instantaneous switch to the new comoving system requires
infinite acceleration and results in a jump of proper time.

But we can try to consider the space trip of Alice without the change of the
reference system at the return point. We compare the situations as seen from
the earth and from the reference system of the spaceship, synchronised at the
starting point.

In the rest frame of the earth the target star of the journey is at a fixed
distance $x_S$, while in the reference system of the spaceship according to
eq.(\ref{lor}) at the state of synchronisation the local time at $x_S$ is
$t_S'=-\gamma\beta x_S$, the position is $x_S'=\gamma x_S$ and the star is
moving towards the spaceship with velocity $-\beta$. The star passes the
spaceship at
\begin{equation}
t_1'=\frac{x_S'}{\beta} +t_S'= \frac{\gamma x_S}{\beta}-\gamma\beta x_S =
\frac{\gamma x_S}{\beta} (1-\beta^2) =\frac{x_S}{\beta\gamma}
\end{equation}
We could, of course, derive the same relation immediately from eq.(\ref{lor})
by setting $t=\beta x_S$ and $x=x_S$. The detour was only to demonstrate the
importance of taking into account missing simultaneity of distant events. In
the comoving system the proper time interval is equal to $t_1'$. In the
earth-bound system we have
\begin{equation}
\Delta\tau = \sqrt{-x_S^2+(x_S/\beta)^2}= \frac{x_S}{\beta\gamma},
\end{equation}
the same value as $t_1'$, as must be expected. The coordinate times are
different, as times at different locations are compared, the proper time is
the same in both systems.

The back journey is a little bit more complicated. Alice in her spaceship
changes the velocity at the turning point. But to keep things simple, we
consider this change only in the system $(x',t')$, as we know that a further
change of the reference system at the turning point would bring in problems
of acceleration and thus of synchronisation. The situation is different now
from the first part of the journey, however. Now for Alice the earth is no
longer at a fixed location, but moving apart at velocity $-\beta$. She is
hunting for a moving target. Thus in her reference system the speed must be
higher than that of the earth. In the non-relativistic case, if the time back
to earth should be equal to $t_1'$, her speed must be $-2\beta$. But if
$\beta > 0.5$, she comes into trouble, as her spaceship is limited to the
speed of light. The back journey will take more time than the first part.

In the limiting case $\beta =1$ the time would even be infinite. No light
pulse can catch up with another pulse, emitted at an earlier time by the same
source. In the relativistic case we must change our problem a little bit. Now
we put the question, when and where will Alice catch up with the moving
earth, when her spaceship now moves with speed $-\beta_2'$, measured in the
reference frame $(x',t')$. The earth is receding at $-\beta$ since $t'=0$,
she herself is moving with $-\beta_2'$ since $t'=t_1'$ Thus she will meet her
twin on earth at $t_2'$, given by
\begin{equation}
-\beta t_2'=-\beta_2'(t_2'-t_1'),
\end{equation}
leading to
\begin{equation}
\label{prime}
 t_2'=\frac{\beta_2'}{\beta_2'-\beta}\;t_1', \qquad
  x_2'=-\frac{\beta \beta_2'}{\beta_2'-\beta}\;t_1'
\end{equation}
and
\begin{equation}
\Delta\tau_2'=\sqrt{1-\beta^2}\frac{\beta_2'}{\beta_2'-\beta}\;t_1'=\frac{\beta_2'}{\beta}
\frac{1-\beta^2}{\beta_2'-\beta}\;x_S.
\end{equation}
If we insert the values of eq.(\ref{prime}) into eq.(\ref{lor}) and solve for
the time interval in rest frame of the earth, we find $\Delta\tau_2\equiv
t_2=\Delta\tau_2'$. There is no change of the proper time between Alice and
Bob.

This simple example shows that it is only the mixing up time intervals,
measured in different reference frames, which leads to contradictions. If
perceived time intervals depend on distance, it is no longer meaningful to
add time intervals measured in different reference systems, which are in
relative motion. The only quantity, which is of physical relevance, is the
space-time distance of events or the proper time interval, which must not
confounded with the perceived time interval measured in some reference frame.
This topic has already been discussed by Kracklauer \cite{krack} in 2001, but
still there are numerous newer papers, which ignore the invariance of proper
time and insist on the existence of the time discrepancy.

\section{Real effects of Lorentz invariance}
Though there exist no local time discrepancies which depend on the course of
the world line between two events, there are several real effects, which are
explained by the Lorentz invariance of the basic laws of physics. But all
these effects are related to our local observations of physical processes,
generated in systems, which are moving with respect to the local rest frame.

One well known example is the apparent increase of life time of unstable
particles like muons, when they approach the earth at velocities close to the
speed of light.

Another is the red shift of light, emitted from moving sources. Though the
light always reaches us with the velocity $c$, the wavelength is shifted.
There is no change of the relative velocity, as we know it from Doppler
shift, but it is the different time scale at emission, which leads to the
wavelength shift at observation.

The difference between Doppler shift and Lorentz shift clearly shows up in
cosmological observations. The observed red shift of light from distant
objects can be explained by a continuous expansion of space, which is
equivalent to a continuous local acceleration or a recession velocity
proportional to distance. This leads to a change of time scale proportional
to distance between emission and observation. This change does not only
affect the frequency of light, but influences all time dependent processes.
One well observed effect is the dilatation of the time scale of distant
supernovas.

But there is a strong caveat in this interpretation of cosmological red shift
and time dilatation. Contrary to Doppler shift, which occurs only with
motions in the direction of observation, the Lorentz shift of time scale is
independent of the direction of acceleration. A continuous acceleration
perpendicular to the direction of observation and acceleration in this
direction will result in exactly the same time shift. If space is curved,
every geodesic motion must be regarded as accelerated. Thus from red shift or
time dilatation measurements we cannot decide, if space is expanding or if
space is curved.

Only independent measurements, which are affected only by the spatial
component of space-time, like the size distribution of distant galaxies or
clusters, can help to decide if the cause of red shift is expansion or
curvature of space. There is one consequence which remains in both cases. If
Lorentz invariance is valid throughout the entire universe, there is no
sensible definition of a global time scale. Time varies with distance. As a
reasonable definition of time to describe distant events in the universe, one
can only use the running time of light with respect to our local reference
frame.

The persistent discussions of the twin paradox demonstrate that we have not
yet fully understood all the consequences of Lorentz invariance of
interactions as well in the regime of mechanics as in gravity. But in the
last decades a huge amount of measurements and observations has been
accumulated, so that we should be able to decide, if Lorentz invariance is
really the governing principle, not only of electromagnetism, but also of
gravity.

\end{document}